\documentclass[11pt]{article}
\usepackage[english]{babel}
\usepackage[a4paper,top=2cm,bottom=2cm,left=2.2cm,right=2.2cm,marginparwidth=1.75cm]{geometry}
\usepackage{setspace}
\usepackage{amsmath}
\usepackage{graphicx}
\usepackage[colorlinks=true, allcolors=blue]{hyperref}

\usepackage[utf8]{inputenc}
\usepackage[T1]{fontenc}
\usepackage[english]{babel}
\usepackage{graphicx}
\usepackage{amsmath,amssymb}
\usepackage[numbers]{natbib}
\usepackage{hyperref}
\usepackage{titlesec}
\usepackage{caption}
\usepackage{setspace}
\usepackage{multicol}
\usepackage{ragged2e} 
\usepackage[dvipsnames]{xcolor} 
\usepackage{tcolorbox} % Paquete para la cajita de color del abstract 
\usepackage{authblk}

\makeatletter

\renewcommand\AB@affilsepx{, }

\def\myinlineaffils{%
  {\small
   \renewcommand{\and}{, }
   \AB@affillist}
}

\renewcommand{\maketitle}{%
  \begin{center}
    {\LARGE\@title\par}
    \vskip 1em
    {\large
      \setlength{\parskip}{0.3em}
      \setlength{\parindent}{0pt}
      \@author
      \par}%
    \vskip 1em
    \vskip 1em
  \end{center}
}
\makeatother

\title{Why the Northern Hemisphere Needs a 30–40 m Telescope and the Science at Stake: Resolved Stellar Populations studies in M31 and its Satellites}

\author[1,2]{C. Gallart}
\author[1,2]{E. Fern\'andez-Alvar}
\author[1,2]{A.B.A. Queiroz}
\author[1,2]{A. Aparicio}
\author[3]{B. Anguiano}
\author[1,2]{G. Battaglia}
\author[1,2]{M. Beasley}
\author[4]{T. Bensby}
\author[5]{G. Bono}
\author[6]{V. Braga}
\author[7]{L. Carigi}
\author[8]{L. Casamiquela}
\author[9,10]{S. Cassisi}
\author[11]{C. Chiappini}
\author[12]{V. P. Debattista}
\author[13]{A. del Pino}
\author[14,15]{I. Escala}
\author[16]{A. M. N. Ferguson}
\author[6]{G. Fiorentino}
\author[15,17]{K. M. Gilbert}
\author[18]{P. Guhathakurta}
\author[19]{R. Ibata}
\author[20]{E. N. Kirby}
\author[21]{K. Kuijken}
\author[22]{S. Larsen}
\author[3]{D. Mart\'\i nez-Delgado}
\author[24]{C. Mart\'\i nez-V\'azquez}
\author[25]{D. Massari}
\author[11]{I. Minchev}
\author[1,9]{M. Monelli}
\author[26]{J. F. Navarro}
\author[27]{M. Ness}
\author[28]{S. Okamoto}
\author[24]{K. Olsen}
\author[30]{S. Ortolani}
\author[31]{P. A. Palicio}
\author[32]{I. P\'erez}
\author[1,2]{F. Pinna}
\author[1,2]{A. Prieto}
\author[33]{J. Read}
\author[31]{A. Recio-Blanco}
\author[34]{M. Rejkuba}
\author[30]{A. Renzini}
\author[35]{R.M. Rich}
\author[32]{T. Ruiz-Lara}
\author[31]{M. Schultheis}
\author[6]{M. Tantalo}
\author[1,2]{G.F. Thomas}
\author[1,2]{A. Vazdekis}
\author[1,2]{E. Villaver}
\author[14]{M. Zoccali}

\affil[1]{Instituto de Astrof\'\i sica de Canarias (IAC), Spain}

\affil[2]{Universidad de La Laguna, Spain}

\affil[3]{Centro de Estudios de F\'\i sica del Cosmos de Arag\'on (CEFCA), Spain}

\affil[4]{Division of Astrophysics, Lund University, Sweden}

\affil[5]{University of Rome, Tor Vergata, Italy}
\affil[6]{INAF—Osservatorio Astronomico di Roma, Italy}

\affil[7]{Instituto de Astronom\'\i a CU, Universidad Nacional Aut\'onoma de México, M\'exico}

\affil[8]{LIRA, Observatoire de Paris, CNRS, France}

\affil[9]{INAF - Astronomical Observatory of Abruzzo, Italy}

\affil[10]{INFN - University of Pisa, Italy}

\affil[11]{Leibniz-Institut f\"ur Astrophysik Potsdam, Germany}

\affil[12]{Jeremiah Horrocks Institute, University of Lancashire, UK}

\affil[13]{Instituto de Astrof\'\i sica de Andaluc\'\i a – Consejo Superior de Investigaciones Científicas (IAA-CSIC), Spain}

\affil[14]{Institute of Astrophysics, Pontificia Universidad Cat\'olica, Chile}

\affil[15]{Space Telescope Science Institute, USA}

\affil[16]{Institute for Astronomy, University of Edinburgh, Royal Observatory, UK}

\affil[17]{Johns Hopkins University, USA}

\affil[18]{University of California Santa Cruz, USA}

\affil[19]{Universit\'e  de Strasbourg, CNRS, Observatoire astronomique de Strasbourg, France}

\affil[20]{University of Notre Dame, USA}

\affil[21]{Leiden Observatory, Leiden University, The Netherlands}

\affil[22]{Department of Astrophysics/IMAPP, Radboud University, The Netherlands}

\affil[24]{NSF NOIRLab, USA}

\affil[25]{INAF - Osservatorio di Astrofisica e Scienza dello Spazio di Bologna, Italy}

\affil[26]{University of Victoria, BC, Canada}

\affil[27]{Research School of Astronomy \& Astrophysics, Australian National University, Australia}

\affil[28]{National Astronomical Observatory of Japan, Japan}

\affil[30]{INAF, Osservatorio Astronomico di Padova, Italy}

\affil[31]{Universit\'e C\^ote d’Azur, Obs. de la C\^ote d’Azur, CNRS, France}

\affil[32]{Universidad de Granada, Departamento de F\'\i sica Te\'orica y del Cosmos, Spain}

\affil[33]{University of Surrey, Physics Department, UK}

\affil[34]{European Southern Observatory (ESO), Germany}

\affil[35]{Department of Physics and Astronomy, UCLA, USA}

\begin{document}
\maketitle

\newpage

\begin{tcolorbox}[colback=RoyalBlue!5!white,colframe=black!75!black, width=\textwidth]
\justifying
{\noindent A 30m class optical/near-IR telescope in the Northern Hemisphere, equipped for diffraction-limited imaging and high-resolution, multi-object spectroscopy of faint stars, would enable a transformational investigation of the formation and evolution of M31 and its satellite system—on par with what Gaia, the HST, and other major photometric and spectroscopic facilities have achieved for the Milky Way (MW) and its satellites. The unprecedented detail obtained for our home system has reshaped our understanding of the assembly of the MW disk, halo, and bulge, and that of its satellites, which now serve as a benchmark for galaxy formation and evolution models. Extending this level of insight to the M31 system —that of the nearest massive spiral and the only one for which such a comprehensive, resolved stellar population study is feasible, will allow us to address a fundamental question: how representative is the MW and its satellite system within the broader context of galaxy evolution? }
\end{tcolorbox}

\section{Introduction and Motivation}
Disk and dwarf galaxies contain a substantial fraction of the stellar mass in the present-day Universe, yet the physical processes that drive their formation and evolution remain only partially understood. Current hydrodynamic cosmological simulations have made remarkable progress in producing realistic disks and reconciling predictions of $\Lambda$CDM paradigm with small scale structure 
but fundamental questions at the heart of galaxy formation and evolution theory persist. Central challenges involve the diverse physical mechanisms that regulate star formation and feedback, the efficiency and radial dependence of stellar migration, the formation of stable disks, the origin and longevity of spiral structure, the origin of thick and thin disks, bulges and bars, as well as the role of mergers, satellite and gas accretion in shaping present day galactic components. These processes are tightly interwoven with the hierarchical growth of structure, and the details remain highly debated. 

Two complementary strategies dominate the study of galaxy formation and evolution, each facing distinct limitations when applied independently. The study of distant galaxies through integrated light reveals broad evolutionary trends across cosmic time but cannot disentangle the complex, multi-generational processes that build individual systems. In contrast, “galactic archaeology” reconstructs the full evolutionary history of individual systems with unparalelled accuracy by analysing their resolved stellar populations. Deep color–magnitude diagrams (CMDs) reaching the oldest main sequence turnoffs (oMSTOs) yield precise ages and time-resolved star-formation histories (SFH); high-resolution spectroscopy provides chemical abundance patterns that trace enrichment processes; finally, chemo-dynamics reveals the imprints of accretion, mergers, and internal evolution.

However, at present, this level of detail is achievable only for the MW and its nearest neighbours within the Local Group. The transformative impact of Gaia and large spectroscopic surveys has revolutionized our understanding of the formation of the Galaxy’s disk, halo, and bulge, which are becoming a critical benchmark for galaxy models. Yet it is key to understand whether the evolution of the MW and its system of satellites is typical. The nearest large spiral galaxy, M31 (Andromeda), and its satellites, is the only external system for which a similarly comprehensive and detailed approach will ever be feasible, and it is only accessible from the ground in the Northern Hemisphere. M31 and the MW are comparable in mass, luminosity, and Hubble type, but they also exhibit striking differences—ranging from disk structure and bulge prominence to the properties of their satellites and halo populations [1]. These contrasts strongly suggest different assembly histories, making M31 an essential laboratory for testing the generality of the evolutionary patterns inferred for the MW. Moreover, Andromeda’s companions, including the smaller disk galaxy M33 and dwarf ellipticals such as M32, NGC 205, and NGC 147, offer uniquely diverse objects in which to examine the interplay between environment, star formation, and chemical evolution.

\section{The needed multi-faceted approach} 
\subsection{Detailed star formation histories} 

The SFH of a galaxy is central to understanding its formation and evolution. It can be defined as the amount of mass transformed into stars, as a function of time and metallicity, necessary to account for its current baryonic content and chemical enrichment. It is the result of, and thus, it allows to identify, major evolutionary events such as mass accretion, interactions, or bar and spiral arm formation. It therefore provides key information to constrain theoretical models of the physical processes that govern the Universe. The most direct way to derive a precise and quantitative SFH is through the analysis of the resolved stars CMD. [2] have shown that the comparison of CMDs reaching the oMSTO  with model CMDs are able to yield extremely precise age-metallicity distributions that are directly linked to spatially resolved SFHs. A large amount of HST and JWST observing time has been used to obtain such CMDs for a few fields in the external parts of M31 [3,4], M33 [5], and in low surface brightness dwarf galaxies at similar distances [6,7]. However, central areas of spirals and denser dwarfs need an order of magnitude higher spatial resolution as well as sensitivity, to obtain the required deep CMDs (down to K$_{AB}$=27.5) at the distance of $\simeq$ 1 Mpc. This is achievable with a diffraction limited (ideally down to $\simeq 1 \mu m$) 30m class telescope [8].

\subsection{Detailed chemodynamics}
The morphological differentiation of galaxies, including the origin of the dwarf galaxy types, results from the combination of hierarchical and secular processes that shape their structure and kinematics. In parallel, star formation transforms the baryonic content and leads to chemical enrichment. The analysis of MW stellar populations has revealed that combining information on their kinematics and chemistry is essential to identify groups of stars formed in different environments [9].

A 30m class telescope providing precise positions, stellar and cluster proper motions, accurate radial velocities and detailed chemical abundances across the M31 system would allow us to replicate many of the studies currently only feasible for the MW, such as identifying chemodynamic substructures in the halo and bulge and characterizing the globular cluster (GC) population. This information would also be essential to characterize the origin and evolution of the disk components that are the defining features of a whole class of galaxies in the Universe.  

For this purpose, the discriminating abundance ratio [$\alpha$/Fe] could be easily measured with a precision of $\sim$0.15 dex in spectra of R$\sim$5000 and S/N$\sim$30 for large samples of RGB stars (K$_{AB}\sim$19.5-21.5), allowing to gather sufficient stellar samples to explore distinct chemical trends. Larger spectral resolution would allow to measure more precise abundances, and is usually required for individual key elements like Na, O and Al, needed to explore the chemical anti-correlations observed in GCs, or n-capture elements like Ba. Access to a wide range of chemical species allows tighter constraints on disk formation models, with a level of detail currently achievable only in the MW.

\section{Is the MW typical? A comparable study of the M31 system}

A 30m class telescope would allow a study of the M31 system of similar scope as the one afforded by Gaia and major ground-based spectroscopic and photometric surveys for the MW and its satellite system. This study has revolutionised the understanding of our immediate neighbourhood and the physical models of galaxy formation and evolution. However, a single template may lead to biased conclusions, and thus, a comparable study of the M31 system, only visible from the North, is essential. Thanks to the spatial resolution and huge light collecting power, the impact of a 30m class telescope would be critical to study the highest surface brightness regions, which would provide a high target density, and thus, the required statistically large samples. In the following, we highlight the scientific questions in which the contribution of such a facility is essential.   

\subsection{The inner region}

The MW is currently the only galaxy in which quantitative resolved stellar population studies of the bulge can be performed. They revealed a superposition of populations and the secular evolution of a bar component and pseudo-bulge, with evidence for merger debris and a spheroidal component [10]. Extending such detailed fossil-record analysis to external bulges is essential for placing the MW bulge in context and for understanding how common such composite structures are. A 30m class telescope in the North would allow us to resolve the bulge of M31, which offers the best opportunity for this comparison, with the added advantage of a lower reddening and contamination by the disk population. M31's bulge shows strong evidence of a composite nature with both a spheroidal component and a bar-related boxy/peanut structure [11], hosting stars with ages ranging from intermediate to old [12]. However, current observing facilities do not offer the necessary depth and resolution. Resolved spectroscopy of individual stars and deep photometry across M31's bulge and inner disc would reveal the ages, metallicities, and kinematics of its subcomponents, clarifying the origin and evolution of M31's inner regions.

\subsection{The globular cluster systems}

GCs are fossil records of the earliest epochs of galaxy assembly. Properly characterizing their ages, chemistry and dynamics is crucial to reconstruct the early events that shaped a galaxy's evolution. The exquisite chrono/chemo/dynamical characterization of MW GCs has enabled the accurate distinction between accreted and in-situ objects. Moreover, the link with the accreted field populations led to the identification of each GC former galaxy progenitor [13] and to a reconstruction of the MW merger tree. Once again, using a single system as template can lead to biased conclusions, and thus a detailed characterization of the M31's cluster population is essential.

In fact, the MW and M31 cluster systems show notable differences. While displaying a relatively wide range of ages, metallicities, and masses, the MW GCs are relatively homogeneous from a structural point of view, while M31 possesses a significant population of luminous and compact GCs and a number of extended GCs with no counterparts in the MW [1]. There are also some indications suggesting that GCs in M31 span a wider age range. The determination of ages, abundances, proper motions and radial velocities of M31's GCs would allow us to understand the origin of these differences and to securely assign clusters to distinct accretion events. 

Additionally, the existence of multi-populations in GCs, whose origin is still not understood, has been a spectacular finding that challenged our view of GCs as prototypes of simple stellar populations. This phenomenon was first identified through light element anti-correlations in RGB stars and then traced to splits in different CMD sequences. Determining the ubiquity of multi-populations by analysing the GC systems of external galaxies is an important step to identify their origin. High resolution spectroscopy of RGB stars in GCs of the M31 system would allow to analyse light element anti-correlations, and provide an important benchmark in the quest for the origin of multi-populations in GCs, specially considering the wider age and mass range of M31's GCs.

\subsection{Disk formation}

The disk is the defining structural component of MW-like galaxies, and revealing its formation and evolution is a key goal of galaxy evolution theory. The MW disk appears to consist of a thick and a thin disk components, broadly characterized by different kinematics and abundance patterns showing intricate relationships between them [14]. Thanks to Gaia and ground based spectroscopy, there has been important progress in the determination of the formation epoch of the MW disk and the characteristics of the chemodynamical duality [14,15,16], which however are difficult to reproduce from a theoretical point of view. The detailed observations required to date the emergence of the disk, and to analyse the (possible) chemical dichotomy with resolved stars are now only possible in the MW, with state-of-the art observations for M31 leading to conflicting conclusions [17,18]. A detailed chemical analysis of M31 would help to understand the ubiquity of this dichotomy and its link with the SFH, the accretion history and the environment of a galaxy. As Gaia, WEAVE, 4MOST and future surveys transform our view of the MW, parallel efforts to obtain detailed spectroscopic coverage of M31 will be crucial for placing both galaxies in a cosmological context.

\subsection{The dE companions and M33}
The M31 satellite system includes galaxy types not present in the MW system. Most notably, it contains dSph galaxies significantly more luminous and higher surface brightness than the most luminous MW dSph satellites and a true dE galaxy, M32, which is possibly the closer counterpart (albeit with drastic differences) of larger elliptical galaxies. These are galaxy types that abound in galaxy clusters, and we have the rare opportunity of study them in detail locally. Finally, M33 offers the opportunity to study a dwarf spiral galaxy and understand how the formation of spiral structure and stellar disks depend on different galactic sizes.

\section*{References}
{\footnotesize
{\bf [1]} Skillman et al. 2017, ApJ, 837:102; 
{\bf [2]} Gallart et al. 2024, A\&A, 687:A168; 
{\bf [3]} Brown et al. 2009, ApJSS, 184:152;
{\bf [4]} Bernard et al. 2018, MNRAS, 420:2625;
{\bf [5]} Barker et al. 2011, MNRAS, 410:504;
{\bf [6]} Gallart et al. 2015, ApJL, 811:L18;
{\bf [7]} Cohen et al. 2025, ApJ, 981:153;
{\bf [8]} Olsen et al. 2003, AJ, 126:452;
{\bf [9]} Horta et al. 2023, MNRAS, 520:5671;
{\bf [10]} Zoccali \& Valenti 2026, In Encycl. of Astrophysics
{\bf [11]} Blaña Díaz et al. 2017, MNRAS, 466:4279;
{\bf [12]} Dong et al. 2018, MNRAS, 478:5379;
{\bf [13]} Massari et al. 2019, A\&A, 630:L4;
{\bf [14]} Queiroz et al. 2020, A\&A, 638:A76;
{\bf [15]} Xiang \& Rix, 2022, Nature, 603:599;
{\bf [16]} Fernández-Alvar et al. 2025, arXiv:2503.19536;
{\bf [17]} Gibson et al. 2025, MNRAS, 542:669;
{\bf [18]} Nidever et al. 2025, In ASS Meeting Abstracts \#243
}
%\begin{multicols}{2}
%\renewcommand{\bibfont}{\footnotesize}  % TamaÃ±o pequeÃ±o
%\bibliographystyle{unsrtnat}     % o el estilo que prefieras
%\bibliography{references_WP}
%\end{multicols}

%\bibliographystyle{unsrturl}
%\bibliography{references}
%\printbibliography

\end{document}